\documentclass[lettersize]{IEEEtran}
\usepackage{array,graphicx,cite,epsfig,amssymb,amsfonts}
\usepackage{bbm,amsmath,epstopdf,algorithm,algorithmic,subfigure, multirow}
\usepackage{comment}
\usepackage{color}
\usepackage{cases}
\usepackage{slashbox}
\newcommand{\be}{\begin{equation}}
\newcommand{\ee}{\end{equation}}
\def\bee#1\eee{\begin{align}#1\end{align}}
\newcommand{\bse}{\begin{subequations}}
\newcommand{\ese}{\end{subequations}}
\newcommand{\nnb}{\nonumber}

\newtheorem{theorem}{\textbf{Theorem}}

\newtheorem{definition}{\textbf{Definition}}

\newtheorem{example}{\textbf{Example}}

\def\ost{\mathsf{offset}}

\newcommand{\bm}[1]{\boldsymbol#1}

\def \ISTR {}

\begin{document}

\title{An Input-Queueing TSN Switching Architecture to Achieve Zero Packet Loss for Timely Traffic}

%\author{Ming~Li$^1$,
%		Lei~Deng$^2$,
%        and~Yunghsiang~S.~Han$^3$
%        \\
%        $^1$Shenzhen University, $^2$Huawei Technology Co. Ltd,  $^3$University of Electronic Science and Technology of China\\
%      %  $^3$Shenzhen Institute for Advanced Study, University of Electronic Science and Technology of China
%}

%\author{\IEEEauthorblockN{Ming Li}
%\IEEEauthorblockA{\textit{College of EIE} \\
%\textit{Shenzhen University}\\
%Shenzhen, China \\
%2019282018@email.szu.edu.cn}
%\and
%\IEEEauthorblockN{Lei Deng}
%\IEEEauthorblockA{\textit{Theory Lab, 2012 Labs} \\
%\textit{Huawei Technology Co. Ltd}\\
%Hong Kong, China \\
%deng.lei2@huawei.com}
%\and
%\IEEEauthorblockN{Yunghsiang S. Han}
%\IEEEauthorblockA{\textit{Shenzhen Institute for Advanced Study} \\
%%\textit{University of Electronic Science and Technology of China}\\
%\textit{UESTC}\\
%Shenzhen, China \\
%yunghsiangh@gmail.com}
%}

\author{ Ming Li, Lei Deng,~\IEEEmembership{Member,~IEEE} and Yunghsiang S. Han,~\IEEEmembership{Fellow,~IEEE} \vspace{-0.3cm}
        % <-this % stops a space
        \thanks{Ming Li is with the Department of Electronic and Electrical Engineering, Southern University of Science and Technology, Shenzhen 518055, China
(email: mingli09117@gmail.com).

Lei Deng is with Theory Lab, 2012 Labs, Huawei Technology Co. Ltd, Hong Kong, China
(email: deng.lei2@huawei.com).

Yunghsiang S. Han is with Shenzhen Institute for Advanced Study, University of Electronic Science and Technology of China, Shenzhen, China
(yunghsiangh@gmail.com).}% <-this % stops a space
}

%\thanks{Ming. Li, L. Deng and S. Han are with
%College of Electronics and Information Engineering, Shenzhen University, Shenzhen 518060, China
%(email: mingli09117@gmail.com, deng.lei2@huawei.com, yunghsiangh@gmail.com).}% <-this % stops a space
%\author{ Ming Li, and Lei Deng,~\IEEEmembership{Member,~IEEE} \vspace{-0.3cm}
%        % <-this % stops a space

%\thanks{Ming. Li and L. Deng are with
%College of Electronics and Information Engineering, Shenzhen University, Shenzhen 518060, China
%(email: 2019282018@email.szu.edu.cn, ldeng@szu.edu.cn).}% <-this % stops a space
%}

\maketitle

\begin{abstract}
Zero packet loss with bounded latency is necessary for many applications, such as industrial control networks, automotive Ethernet, and aircraft communication systems.
Traditional networks cannot meet the such strict requirement, and thus Time-Sensitive Networking (TSN) emerges.
TSN is a set of standards proposed by  IEEE 802 for providing deterministic connectivity in terms of low packet loss, low packet delay variation, and guaranteed packet transport.
However, to our knowledge, few existing TSN solutions can deterministically achieve zero packet loss with bounded latency.
This paper fills in this blank by proposing a novel input-queueing TSN switching architecture, under which
we design a TDMA-like scheduling policy (called M-TDMA) along with a sufficient condition and an EDF-like scheduling policy (called M-EDF) along with a different sufficient condition
to achieve zero packet loss with bounded latency.
\end{abstract}


\section{Introduction}

\IEEEPARstart{M}{any} real-time applications require deterministic services
in terms of deterministic packet delay/latency,\footnote{In the paper, we use delay and latency interchangeably, both of which
refers to a time duration, while deadline refers to a time instance.} deterministic packet loss and/or deterministic packet delay jitter.
Zero packet loss with bounded latency is one of the most challenging goals. Still, it is a must for many applications such as industrial control networks, automotive Ethernet, and aircraft communication systems.
In such applications, all generated packets need to be delivered successfully in a bounded delay without any loss to ensure extreme safety.

Traditional networks can only provide best-effort (BE) service and thus cannot guarantee deterministic quality.
Existing Fieldbus solutions, such as PROFIBUS and CAN, and industrial-Ethernet solutions, such as PROFINET and EtherCAT, can
provide deterministic quality but encounter interoperation issue because they adopt different sets of standards \cite{decotignie2009many,industrialcom}.
To address this issue, IEEE 802 designed Time-Sensitive Networking (TSN) standards \cite{tsnwc} mainly working in layer 2, and IETF proposed Deterministic Networking (DetNet) standards \cite{detnetwc}
primarily working in layer 3, both of which aim at offering a unified solution over the widely used Ethernet standards. In this work, we investigate layer-2 switching solutions and thus pay attention to TSN.
TSN is a set of standards proposed by  IEEE 802 for providing deterministic connectivity in terms of low packet loss, low packet delay variation, and guaranteed packet transport.
TSN has been envisioned as a future solution in industrial communication, and automation systems \cite{tsndefin,industrialcom}.

However, to our knowledge, few existing TSN solutions can deterministically achieve zero packet loss with bounded latency.
First, some leading telecommunication and industrial-automation giants have already marketed TSN switch products,
such as Cisco's IE 4000 series \cite{ciscoswitch} and Moxa's TSN-G5008 series \cite{moxoswitch}.
However, according to the public data sheets of these products, they have only implemented part of TSN standards and have not mentioned that they can achieve
zero packet loss with bounded latency.
Second, in the IEEE 802 TSN standards, bounded latency can be guaranteed by 802.1Qav, 802.1Qbu, 802.1Qbv, 802.1Qch, and 802.1Qcr, etc.,
and ultra-reliability can be secured by 802.1CB, 802.1Qca, and 802.1Qci, etc. However, such standards cannot meet the extreme requirement of zero packet loss with bounded
latency. Finally, even in the TSN research community, we only find very few research papers on achieving zero packet loss with bounded latency.
In particular, \cite{deng2018delay} proposed a throughput-optimal switching scheduling policy for the particular frame-synchronized delay-constrained traffic pattern,
where statistically zero packet loss can be achieved by setting the required timely throughput $R_{i,j}=1/T$ in \eqref{equ:thm-M-EDF-con2}, where $T$ is the common arrival period and hard delay.

This paper fills this blank by proposing an input-queueing TSN switching architecture to deterministically achieve zero packet loss with bounded latency for
time-sensitive (TS) flows. Specifically, we propose two different scheduling policies along with two corresponding sufficient conditions to achieve this goal.
The first one is based on the TDMA (Time Division Multiple Access) scheduling policy, which works for TS flows with arbitrarily different start times.
The second one is based on the EDF (Earliest-Deadline First) \cite{Liu1973} scheduling policy, originally designed for single-processor task execution.
We extend EDF from the many-to-one task-executing system to our many-to-many switching system. We require that part of TS flows start simultaneously while the rest of TS flows can begin at any time.
These two sufficient conditions do not have an inclusion relationship and work for different settings.

\begin{figure*}[t]
   \centering
    \includegraphics[width=0.87\linewidth]{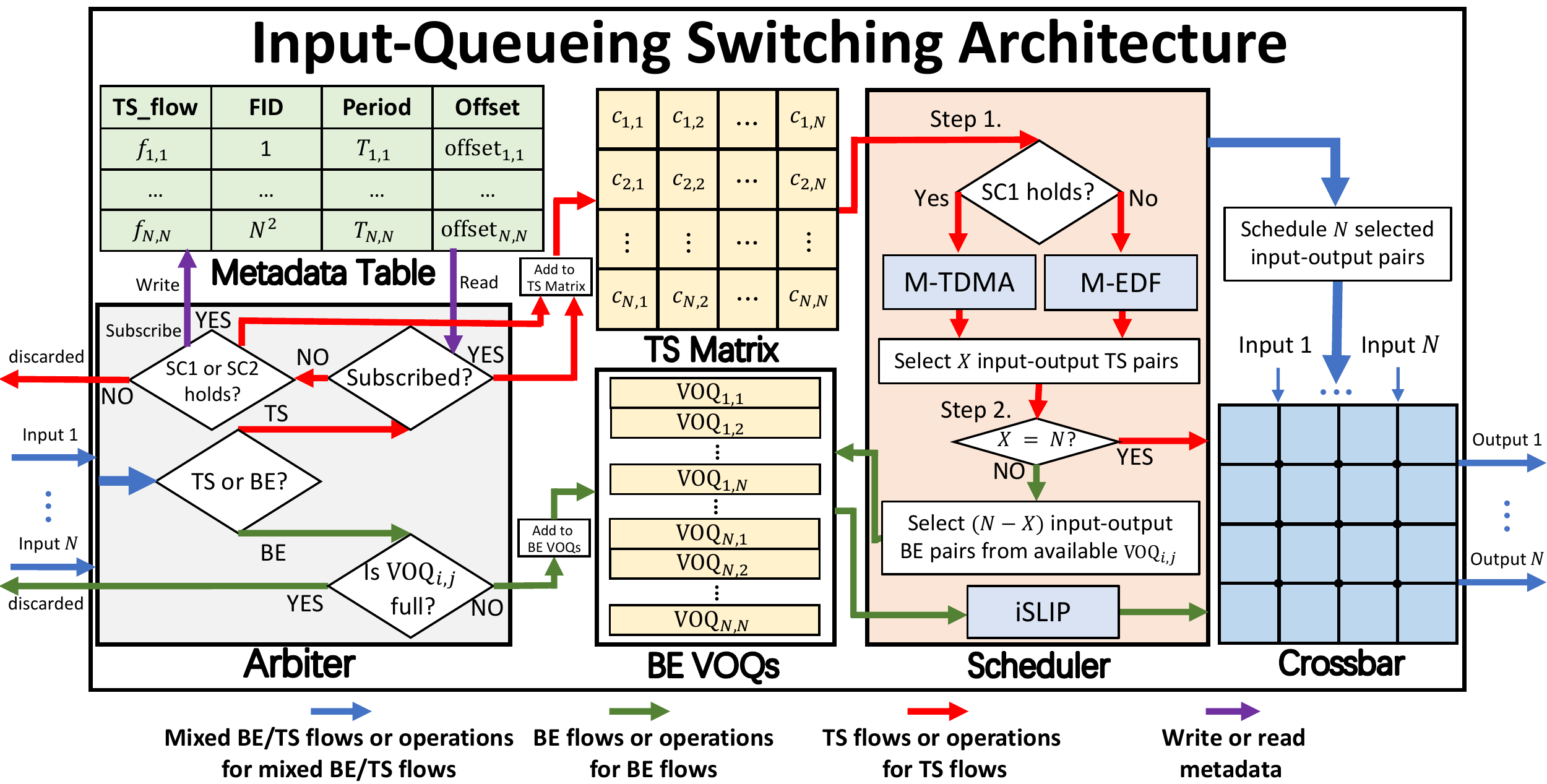}
    \caption{Our proposed input-queueing TSN switching architecture to achieve zero packet loss with bounded latency. \label{fig:switch} }
\end{figure*}

In the rest of this paper, we use $[N]$ to denote set $\{1,2,\cdots,N\}$ for any positive integer $N$.
\ifx \ISTR \undefined
Due to the page limit, some detailed proofs  are deferred to our technical report \cite{TR}.
\else
\fi
%Due to the page limit, some detailed proofs  are deferred to our technical report \cite{TR}.

\section{Our Proposed Input-Queueing Switching Architecture}
In this section, we describe our proposed input-queueing switching architecture to achieve zero packet loss for timely traffic, as shown in
Fig.~\ref{fig:switch}. Before that, we first describe some assumptions. If a data flow has a hard delay constraint,
we call it a time-sensitive (TS) flow; otherwise, we call it a best-effort (BE) flow.
We further assume that any TS flow is periodic and the arrival period is equal to its hard delay, which well models practical scenarios \cite{tsndefin,industrialcom,deng2018delay}.
In addition, we assume that arriving packets of any TS or BE flow are of fixed and equal length, called
cells \cite{mckeown1999achieving}. Each cell needs one slot to be delivered from its input to its output in the switch.
We now detail the six critical components of Fig.~\ref{fig:switch}.

\textbf{Metadata Table.}
A TS flow from input port $i$ to output port $j$ is called TS flow $f_{i,j}$.
For any TS flow $f_{i,j}$, we associate three metadata fields: Flow ID (FID), Period $T_{i,j}$, and Offset $\ost_{i,j}$.
The Metadata Table in Fig.~\ref{fig:switch} records all metadata fields of all TS flows.

\textbf{TS Matrix.} Since we require that the arrival period of any TS flow is equal to its hard delay,
at most one non-expired cell for any TS flow exists in the switch at any slot. Then, we
can construct a binary $N \times N$ matrix to indicate whether there is a non-expired cell for each TS flow,
which is called the \emph{TS Matrix}. If TS flow $f_{i,j}$ has a non-expired cell,
then the entry in the $i$-th row and $j$-th column, denoted by $c_{i,j}$, is 1; otherwise, $c_{i,j}= 0$.
Of course, we should allocate a memory unit to store each non-expired cell. Without introducing ambiguity,
we also call that the TS matrix stores all non-expired cells.

\textbf{BE VOQs.} For any BE flow, we create a virtual output queue (VOQ), which is a first-in-first-out (FIFO) queue, to store its cells. Different from TS flows,
a BE flow could have multiple cells simultaneously in the system, since BE cells will not expire
and thus will be kept in the system forever till its delivery. The VOQ to store the cells of
the BE flow from input $i$ to output $j$ is called VOQ$_{i,j}$. We have in total $N^2$ VOQs
in the switch, as shown in Fig.~\ref{fig:switch}. Note that this is the classical input-queued VOQ switch architecture \cite{mckeown1999islip,mckeown1999achieving}.

\textbf{Arbiter.} To achieve zero packet loss for all TS flows, we need to reduce the contention level. Thus, not all TS flows can be allowed to enter the switch.
This decision component is called the \emph{arbiter}. Specifically, the arbiter first determine whether each arriving flow from input $i$ to output $j$ is a TS flow or a BE flow.
If it is a TS flow (called $f_{i,j}$), the arbiter will look up the Metadata Table to see if it has already been subscribed (i.e., it has been recorded in the Metadata Table).
If it has already been subscribed, the incoming cell of flow $f_{i,j}$ is directly stored in the TS matrix; otherwise, the arbiter
 judges if Sufficient Condition 1 (SC1) as shown in Theorem~\ref{thm:M-TDMA}  or Sufficient Condition 2 (SC2) as shown in Theorem~\ref{thm:M-EDF} holds when we
add $f_{i,j}$ into the Metadata Table.
If either SC1 or SC2 holds, $f_{i,j}$ is subscribed, i.e., it is added to the Metadata Table
otherwise, $f_{i,j}$ is rejected, and the incoming cell is discarded.
If this flow is a BE flow (called $f^{\textsf{BE}}_{i,j}$), the arbiter  checks if VOQ$_{i,j}$ is full.
If VOQ$_{i,j}$ is full, the incoming cell is discarded (i.e., an overflow occurs); otherwise, it is added to VOQ$_{i,j}$.

\textbf{Scheduler.} The scheduler determines which TS flows and which BE flows are scheduled at any slot.
TS flows have higher priorities than BE flows. Thus, the scheduler has two steps. In step 1,
it selects TS flows. From the arbiter, we know that all subscribed TS flows in the metadata table must satisfy either SC1 or SC2.
If SC1 holds, the scheduler runs the M-TDMA policy in Sec.~\ref{subsec:M-TDMA} to
select some TS flows at this slot. Otherwise, if SC2 holds,
the scheduler runs the M-EDF policy in Sec.~\ref{subsec:M-EDF} to select some TS flows at this slot.
If TS flow $f_{i,j}$ is selected but $c_{i,j}=0$ in the TS matrix, $f_{i,j}$ is not selected as it does not have a non-expired cell.
In step 2, the scheduler selects BE flows by running the well-known iSLIP policy \cite{mckeown1999islip}. If TS flow $f_{i,j}$ is selected at this slot in step 1,
then any BE flow $f_{i,k} (\forall k)$ and $f_{k',j} (\forall k')$ cannot be selected in step 2. Namely,
step 2 only utilizes the rest of the available ports to select BE flows without interfering with all selected TS flows in step 1.
If $X$ TS pairs are selected in step 1, then $N-X$ BE pairs are selected in step 2.
By convention, we require that iSLIP must schedule $N-X$ BE pairs even if some BE pairs have empty VOQs.
%After that, the switch schedules all $N$ selected input-output TS and BE pairs at this slot.

\textbf{Crossbar.} All $N$ input-output pairs selected by the scheduler, including $X$ TS flows and $N-X$ BE flows,
will be switched by the crossbar fabric. The delivery in this slot ends.

\section{Two Sufficient Conditions To Achieve Zero Packet Loss}
We consider an $N\times N$ input-queued switch whose architecture is shown in Fig.~\ref{fig:switch} where $N \ge 2$ is a positive integer.
Time is slotted starting from slot 0. Since TS flows have higher priorities than BE flows, and our goal is to achieve zero packet loss for TS flows,
we only consider TS flows in this section.
Later in Appendix D,
we show an example of scheduling both TS flows and BE flows.
%Later in Appendix \ref{app:schedule-both-TS-and-BE-flows},
%we show an example of scheduling both TS flows and BE flows.
Recall that a TS flow from input port $i$ to output port $j$, called flow $f_{i,j}$,
is characterized
its offset $\ost_{i,j}$ and its period $T_{i,j}$, where $\ost_{i,j} \ge 0 $ and $T_{i,j} \ge 1$ are both integers.
Starting from slot $\ost_{i,j}$, flow $f_{i,j}$ has a new cell arrival every $T_{i,j}$ slots.
All flow cells are of the same size and need one slot to be delivered from its input to its output.
In addition, each cell of flow $f_{i,j}$ has a hard delay of $T_{i,j}$ slots. Namely,
if it cannot be scheduled in $T_{i,j}$ slots after its arrival, it will be discarded from the switch.
To achieve zero packet loss, we should ensure that no cell will be discarded. We remark that no TS flow may exist from input $i$ to output $j$. In the rest of this paper, we still call it flow $f_{i,j}$ and
set $\ost_{i,j}=\infty$ and $T_{i,j}=\infty$ by convention.

Due to the physical limitation of crossbar fabric, in each slot,
each input port can transmit at most one cell to one output port,
and each output port can receive at most one cell from one input port.
This is called the \emph{crossbar constraint}.
Following \cite{deng2018delay}, we can equivalently describe the crossbar constraint
by a matching characterized by a matrix\footnote{For simplicity, we call it matching $\bm{M}$, instead of matching matrix $\bm{M}$.}
$\bm{M}=(m_{i,j} \in \{0,1\}: i,j\in [N])$ of an $N$-by-$N$  bipartite graph (see \cite[Fig.~1(b)]{deng2018delay}),
where $m_{i,j}=1$ means that flow $f_{i,j}$ is selected/scheduled and $m_{i,j}=0$ means that flow $f_{i,j}$ is not selected/scheduled.
We also call that matching $\bm{M}$ contains flow $f_{i,j}$ if $m_{i,j}=1$. Alternatively, we also call that flow $f_{i,j}$ is in matching $\bm{M}$, denoted by $f_{i,j} \in \bm{M}$ with a slight abuse of notation.
A binary square matrix is a matching if and only if each row or column contains at most one 1.

Thus, in each slot, the scheduler of Fig.~\ref{fig:switch} needs to select a matching $\bm{M} \in \mathcal{M}$, where $\mathcal{M}$ is the set of all matchings.
In addition, a perfect matching $\bm{M}$ is a matching with $N$ 1's, i.e., $\sum_{i=1}^N \sum_{j=1}^N m_{i,j}=N$. The set of all perfect matchings is denoted by $\mathcal{M}^p \subset \mathcal{M}$.
It is also easy to see that a binary square matrix is a perfect matching if and only if each row or column contains exactly one 1. Next we first
give the definition for \emph{flow decomposition sets} and then propose our two scheduling policies along with their corresponding sufficient conditions to achieve zero packet loss for all TS flows.

\subsection{Flow Decomposition Set} \label{subsec:flow-decomp-set}
A set of $N$ perfect matchings, denoted by $\mathcal{D}=\{\bm{M}_1, \bm{M}_2, \cdots, \bm{M}_N\}$, is called a \emph{flow decomposition set} if the result of sum of all $M_i \in \mathcal{D} \subset \mathcal{M}^p$ is an $N \times N$ all-one matrix, i.e., $\bm{M}_i \in \mathcal{D} \subset \mathcal{M}^p, \forall i \in [N]$, and
\be
\sum_{i=1}^N \bm{M}_i = \bm{1}_{N\times N}. \label{equ:def-D}
\ee
The right-hand side of \eqref{equ:def-D} is an $N \times N$ all-one matrix, and the addition operation of the left-hand side of \eqref{equ:def-D} is in the real number field.
%where the bit-wise sum of all $\bm{M}_i \in \mathcal{D}$ is an $N \times N$ all-one matrix.
%$\bm{1}_{N\times N}$ of the right-hand  side is an $N \times N$ all-one matrix, and the addition operation of the left-hand side is in the real field
Constraint \eqref{equ:def-D} has two properties. First, any flow in a perfect matching $\bm{M}_i \in \mathcal{D}$
is different from any flow in another perfect matching $\bm{M}_j \in \mathcal{D}$ with $j \neq i$.
Second, any flow in the system (within total $N^2$ flows) exists in a matching of set $\mathcal{D}$.
In other words, if we schedule all matchings in $\mathcal{D}$ once, any flow in the system is scheduled precisely once.
Since the set is unordered, it does not matter which perfect matching is called $\bm{M}_i$ in $\mathcal{D}$.
To facilitate further analysis, for any flow decomposition set $\mathcal{D}$, we require that
\emph{ the perfect matching in $\mathcal{D}$, which contains flow $f_{1,i}$,
is called $\bm{M}_i$ for any $i \in [N]$.}

In addition, for ease of presentation, with a slight abuse of matrix notation, we define a \emph{perfect-matching matrix} $\bm{L}$ as an $N$-by-$N$ matrix
%whose entry in the $i$-th row and $j$-th column, denoted by $l_{i,j}$,
whose entry in the $i$-th row and $j$-th column
is a perfect matching in  $\mathcal{M}^p$ containing flow $f_{i,j}$.

For any flow decomposition set $\mathcal{D}=\{\bm{M}_1, \bm{M}_2, \cdots, \bm{M}_N\}$, we can construct a unique perfect-matching matrix
whose entry is required to be a perfect matching in $\mathcal{D}$, denoted by $\bm{L}(\mathcal{D})$. 
For example, the flow decomposition set represented by (12) in Appendix A
yields to the following perfect-matching matrix,
%For example, the flow decomposition set represented by \eqref{equ:app-M1-M4} in Appendix~\ref{app:ex-flow-decomposition-set}
%yields to the following perfect-matching matrix,
% seems to have some problem here. (punctuation mark)
\be
\resizebox{.88\hsize}{!}{$
\bm{L}(\mathcal{D}) = (l_{i,j}: i,j \in [N])=
\left(
  \begin{matrix}
    \bm{M}_1 & \bm{M}_2 & \bm{M}_3 & \bm{M}_4 \\
    \bm{M}_4 & \bm{M}_1 & \bm{M}_2 & \bm{M}_3 \\
    \bm{M}_3 & \bm{M}_4 & \bm{M}_1 & \bm{M}_2 \\
    \bm{M}_2 & \bm{M}_3 & \bm{M}_4 & \bm{M}_1 \\
  \end{matrix}
\right).
$}
\label{equ:ex-L-D}
\ee
Later in Sec.~\ref{subsec:M-TDMA} and Sec.~\ref{subsec:M-EDF}, we will construct a scheduling policy according to $\bm{L}(\mathcal{D})$: in each slot, the scheduler of Fig.~\ref{fig:switch} selects an $\bm{M}_i$ and schedules the flows $f_{i,j}$ according to the positions of  $\bm{M}_i$ in $\bm{L}(\mathcal{D})$. 
Clearly, the first row of $\bm{L}(\mathcal{D})$ must be  $(\bm{M}_1,\bm{M}_2,\cdots, \bm{M}_N)$ and
each row or column of  $\bm{L}(\mathcal{D})$  contains all $\bm{M_}i$'s for $i \in [N]$. Following \cite{mckay2005number}, we know that $\bm{L}(\mathcal{D})$ is a Latin square\footnote{Recall that
a Latin square of order $N$ is an $N \times N$ square matrix containing a set of $N$ different symbols/numbers in every row and column.}
where the underlying symbol set is $\mathcal{D}=\{\bm{M}_1, \bm{M}_2, \cdots, \bm{M}_N\}$.
On the contrary, we can also construct a unique flow decomposition set $\mathcal{D}(\bm{L})$ for any
Latin square $\bm{L}$ of order $N$ whose first row is fixed as $(\bm{M}_1,\bm{M}_2, \cdots, \bm{M}_N)$. %, as shown in the following theorem.

\begin{theorem} \label{the:iff-conditon}
For any $N \times N$ switch, there exists a bijection between the set of all its flow decomposition sets and  the set of all Latin squares of order $N$
whose first rows are fixed as $(\bm{M}_1,\bm{M}_2, \cdots, \bm{M}_N)$.
%A perfect-matching matrix $\bm{L}$ yields to a unique flow decomposition set $\mathcal{D}$ if and only if $\bm{L}$ is a Latin square.
\end{theorem}
\begin{IEEEproof}
Please see Appendix B.
%Please see Appendix~\ref{app:proof-of-the-iff-condition}.
\end{IEEEproof}

Therefore, the total number of flow decomposition sets is equal to the total number
of Latin squares of order $N$ whose first rows are fixed as $(\bm{M}_1,\bm{M}_2, \cdots, \bm{M}_N)$.
In the research area of Latin squares, we can count the total number of Latin squares
and the total number of reduced-form Latin squares (whose first row and the first column are both in the natural order)
for up to $N=11$. It is straightforward to obtain that the total number of Latin squares of order $N$ whose first row is fixed as $(\bm{M}_1,\bm{M}_2, \cdots, \bm{M}_N)$ is equal to $(N-1)!$ times the total number of reduced-form Latin squares of order $N$. Therefore, based on \cite{mckay2005number},
we can count the total number of $\bm{L}(\mathcal{D})$ for an $N \times N$ switch, denoted by $\alpha(N)$ for up to $N=11$.
Since the TSN switches are generally small, we list $\alpha(N)$ for up to $N=7$ in Table~\ref{tab:count}. 

\begin{table}[t]
\centering
\caption{The total number of flow decomposition sets, $\alpha(N)$, for $N=1,2,\cdots,7$ \label{tab:count}}
\begin{tabular}{|c|c|c|c|c|c|c|}
  \hline
  $N$ & 2 & 3 & 4 & 5 & 6 & 7 \\
  \hline
  $\alpha(N)$  & 1 & 2 & 24 & 1,344 & 1,128,960 & 12,198,297,600 \\
 % $\alpha(N)$  & 1 & 2 & 24 & 1,344 & 1,128,960 & $\approx 1.2 \times 10^{10}$ & $\approx 2.7 \times 10^{15}$ \\
\hline
\end{tabular}
\end{table}

\subsection{Sufficient Condition 1 and Matching-based TDMA (M-TDMA) Scheduling Policy} \label{subsec:M-TDMA}
We now propose our Sufficient Condition 1 (SC1) and its corresponding scheduling policy to achieve zero packet loss for timely traffic,
as shown in Theorem~\ref{thm:M-TDMA} shortly.

First, for any flow decomposition set $\mathcal{D}=\{\bm{M}_1,\bm{M}_2,\cdots,\bm{M}_N\}$, we define a matching-based TDMA (M-TDMA) scheduling policy as follow\footnote{In fact, we can schedule such $N$ perfect matchings in any permutation order as long as the order is repeated every $N$ slots.}:  $\forall q=0,1,\cdots$
\begin{itemize}
\item Schedule $\bm{M}_1$ at slots $qN$, 
\item Schedule $\bm{M}_2$ at slots $qN+1$,
\item $\cdots$
\item Schedule $\bm{M}_N$ at slots $qN+(N-1)$, 
\end{itemize}

where the M-TDMA scheduling policy starts at slot 0.
Note that every $N$ slots consist of a scheduling period and $q$
is the index of scheduling periods. Matchings $(\bm{M}_1,\cdots,\bm{M}_N)$
are repeatedly scheduled in order in every scheduling period. That is why we
call it a matching-based TDMA  policy.

\begin{theorem}[Sufficient Condition 1 (SC1)] \label{thm:M-TDMA}
For any $N\times N$ switch with TS traffic offset matrix  $\boldsymbol{\ost}=(\ost_{i,j}: i,j \in [N])$ and traffic period matrix $\boldsymbol{T}=(T_{i,j}: i,j\in[N])$,
if
\be
T_{i, j} \ge N, \quad \forall i, j \in [N], \label{equ:sc1}
\ee
then the M-TDMA policy for any flow decomposition set $\mathcal{D}=\{\bm{M}_1,\bm{M}_2,\cdots,\bm{M}_N\}$  achieves zero packet loss.
\end{theorem}
\begin{IEEEproof}
We give a proof sketch here and defer the detailed proof to
Appendix E in our.
%\ifx \ISTR \undefined
%Appendix~\ref{app:proof-of-thm-M-TDMA}.
%\else
%Appendix~\ref{app:proof-of-thm-M-TDMA}.
%\fi

M-TDMA policy guarantees that each flow is scheduled within $N$ slots staring from any slot.
Thus, any packet of any flow $f_{i,j}$ will be scheduled within $N$ slots after its arrival,
which happens before its expiration and the next packet's arrival of this flow due to \eqref{equ:sc1}.
Thus, we achieve zero packet loss for all timely flows.
\end{IEEEproof}

\subsection{Sufficient Condition 2 and Matching-based EDF (M-EDF) Scheduling Policy} \label{subsec:M-EDF}
M-TDMA scheduling policy only works for the situation of $T_{i,j} \ge N, \forall i,j \in [N]$. It cannot guarantee
zero packet loss if there exists a TS flow $f_{i,j}$ with period $T_{i,j} < N$. This does not mean that no policy
can achieve zero packet loss if some $T_{i,j} < N$. We now describe a different sufficient condition
on scheduling policy to address this issue.
We first define a $T$-vector as follows.
\begin{definition}
For any positive integer $N$, a $T$-vector is a 1-by-$N$ vector, denoted by $\overrightarrow{T}=(T_1,T_2,\cdots,T_N)$,
if any $T_k$ is a positive integer and
\be
\sum_{k=1}^N \frac{1}{T_k} \le 1. \label{equ:condition-basis}
\ee
\end{definition}

For any $T$-vector $\overrightarrow{T}=(T_1,T_2,\cdots,T_N)$, we construct a
\emph{virtual} single-processor task scheduling system \cite{Liu1973}. Specifically,
a single processor needs to execute $N$ periodic delay-constrained tasks where
task $k$ has period $T_k$ for all $k \in [N]$. Starting from
slot 0, task $k$ generates a request every $T_k$ slots. The processor
needs one slot to complete a request, and each request at task $k$ has a hard delay of $T_k$ slots.
Thus,  task $k$ is analogous to the timely traffic with zero offset and period $T_k$.
Liu and Layland in \cite{Liu1973} proposed the famous
earliest-deadline first (EDF) scheduling policy and proved that
EDF completes  all requests (equivalent to achieve zero packet loss) if $\overrightarrow{T}=(T_1,T_2,\cdots,T_N)$ is a $T$-vector (see \cite[Theorem 7]{Liu1973}).
At each slot, EDF schedules the request of all $N$ tasks, which has the earliest deadline and breaks ties arbitrarily.
We now define $a^{\textsf{EDF}}_t(\overrightarrow{T}) \in [N]$ as the index of the scheduled task at slot $t$
when EDF schedules all $N$ tasks spanned by $T$-vector $\overrightarrow{T}$.
An example of $\left(a^{\textsf{EDF}}_t(\overrightarrow{T}):t=0,1,2,\cdots \right)$ sequence is shown
in Appendix C.
%An example of $\left(a^{\textsf{EDF}}_t(\overrightarrow{T}):t=0,1,2,\cdots \right)$ sequence is shown
%in Appendix~\ref{app:an-example-EDF}.

Now we propose  Sufficient Condition 2 (SC2) and
the corresponding matching-based EDF (M-EDF) scheduling policy.

\begin{theorem}[Sufficient Condition 2 (SC2)] \label{thm:M-EDF}
For any $N\times N$ switch with TS traffic offset matrix  $\boldsymbol{\ost}=(\ost_{i,j}: i,j \in [N])$ and traffic period matrix $\boldsymbol{T}=(T_{i,j}: i,j \in [N])$,
if there exists a flow decomposition set $\mathcal{D}=\{\bm{M}_1,\bm{M}_2,\cdots, \bm{M}_N\}$ and a $T$-vector $\overrightarrow{T}=(T_k:k \in [N])$, where $T_k$ is the corresponding scheduling period of $\bm{M}_k$, such that for any flow $f_{i,j} \in \bm{M}_k$,
% if there exists a $T$-vector $\overrightarrow{T}=(T_1, T_2, \cdots, T_N)$ and a flow decomposition set $\mathcal{D}=\{\bm{M}_1,\bm{M}_2,\cdots, \bm{M}_N\}$ such that for any flow $f_{i,j} \in \bm{M}_k$,
either
\be
\quad T_{i, j}= T_k \text{ and } \ost_{i,j}=0 \label{equ:thm-M-EDF-con1}
\ee
or
\be
\quad T_{i, j} \ge 2 T_k -1 \text{ and } \ost_{i,j} \ge 0 \label{equ:thm-M-EDF-con2}
\ee
holds,
then the M-EDF policy, which schedules matching $\bm{M}_{a^{\textsf{EDF}}_t(\overrightarrow{T})}$ at any slot $t$,
achieves zero packet loss.
\end{theorem}

\begin{IEEEproof}
We give a proof sketch here and defer the full proof to
Appendix F.
%\ifx \ISTR \undefined
%Appendix~\ref{app:proof-of-thm-M-EDF} in our technique report \cite{TR}.
%\else
%Appendix~\ref{app:proof-of-thm-M-EDF}.
%\fi
%Appendix~\ref{app:proof-of-thm-M-EDF} in our technique report \cite{TR}.

If condition \eqref{equ:thm-M-EDF-con1} holds, the $s$-th packet of flow $f_{i,j}$ in the real switching system arrives and expires
at exactly the same time as the $s$-th request of task $k$ in the \emph{virtual} single-processor task scheduling system. Since EDF achieves
zero packet loss for any $T$-vector, M-EDF also achieves zero packet loss for all timely flows.

If condition \eqref{equ:thm-M-EDF-con2} holds,
M-EDF policy guarantees that flow $f_{i,j}$ is scheduled once within $2 T_k -1$ slots staring from any slot.
Thus, any packet of any flow $f_{i,j}$ will be scheduled within $2 T_k -1$ slots after its arrival,
which happens before its expiration and the next packet's arrival of this flow due to \eqref{equ:thm-M-EDF-con2}.
Thus, we achieve zero packet loss for all timely flows.
\end{IEEEproof}

We remark that Sufficient Condition 2 in Theorem~\ref{thm:M-EDF} is more difficult to check than Sufficient Condition 1 in Theorem~\ref{thm:M-TDMA}.
For any offset matrix $\bm{\ost}$ and any period matrix $\bm{T}$,
we need to find a $T$-vector and flow decomposition set $\mathcal{D}$ to satisfy \eqref{equ:thm-M-EDF-con1} or \eqref{equ:thm-M-EDF-con2}.
In this paper, we propose a brute-force algorithm to find such a $T$-vector and flow decomposition set $\mathcal{D}$,
which works for small $N$, e.g., when $N \le 6$. Following Theorem~\ref{the:iff-conditon}, we first enumerate all flow decomposition sets based on enumeration of all Latin squares of order $N$ whose first rows
are fixed as $(\bm{M}_1,\bm{M}_2, \cdots, \bm{M}_N)$.
For any perfect matching $\bm{M}_k$ in each flow decomposition set $\mathcal{D}=\{\bm{M}_1, \bm{M}_2, \cdots, \bm{M}_N\}$,
we set a $T_k$ value as follows.
We first define two values,
\be
t_1 = \min_{ f_{i,j} \in \bm{M}_k: i,j \in [N], \ost_{i,j}=0} T_{i,j}, \label{equ:t1}
\ee
which is set to be $\infty$ if $\ost_{i,j}>0$ for all $i,j \in [N]$, and
\be
t_2 = \min_{ f_{i,j} \in \bm{M}_k: i,j \in [N]} \left \lfloor \frac{T_{i,j}+1}{2} \right \rfloor. \label{equ:t2}
\ee
We can see that $t_1 \ge t_2$. It is also easy to verify that the maximum $T_k$, such that
either condition \eqref{equ:thm-M-EDF-con1} or condition \eqref{equ:thm-M-EDF-con2} for any flow $f_{i,j} \in \bm{M}_k$, is either $t_1$ or $t_2$.
Clearly, if we set $T_k=t_2$, any flow $f_{i,j} \in \bm{M}_k$ satisfies  condition \eqref{equ:thm-M-EDF-con2}. Now we only need
to check if we can set $T_k=t_1$. We need to consider each flow $f_{i,j} \in \bm{M}_k$ and check whether condition \eqref{equ:thm-M-EDF-con1} or condition \eqref{equ:thm-M-EDF-con2} holds
when $T_k=t_1$. If so, we set $T_k=t_1$; otherwise, we set $T_k=t_2$.

After that, we construct a vector $\overrightarrow{T}=(T_1,T_2,\cdots, T_N)$. We then check if it is a $T$-vector based on \eqref{equ:condition-basis}.
If so, we have found such  a $T$-vector $\overrightarrow{T}=(T_1,T_2,\cdots, T_N)$
and a flow decomposition set $\mathcal{D}$ to meet   Sufficient Condition 2 in Theorem~\ref{thm:M-EDF}. If we cannot find
such a $T$-vector to meet \eqref{equ:condition-basis} for all flow decomposition sets,
the given offset matrix $\bm{\ost}$ and period matrix $\bm{T}$ cannot meet Sufficient Condition 2 in Theorem~\ref{thm:M-EDF}.
How to find more efficient algorithms to check sufficient condition 2 for large $N$ is left as a future direction.

\section{Examples}
In the previous section, we have proposed two sufficient conditions. However, we cannot say that one is more strict than the other.
In this section, we will illustrate two examples to justify this argument. The feasibility of these two examples has been verified by the computer simulation.

\begin{example} \label{ex1:TDMA}
Consider a $4 \times 4$ switch with TS traffic whose offset matrix and period matrix are respectively set as follows,
\bee
\resizebox{.88\hsize}{!}{$
\bm{\ost}=
\left(
  \begin{array}{cccc}
    1 & 2 & 3 & 1 \\
    2 & 0 & 5 & 2 \\
    3 & 4 & 2 & 4 \\
    8 & 2 & 3 & 3 \\
  \end{array}
\right),
\bm{T}=
\left(
  \begin{array}{cccc}
    4 & 4 & 5 & 5 \\
    5 & 5 & 5 & 5 \\
    4 & 4 & 4 & 6 \\
    4 & 5 & 5 & 5 \\
  \end{array}
\right).
$}
\label{equ:ex1}
\eee
Since $T_{i,j} \ge N=4$ for all flow $f_{i,j}$, \eqref{equ:ex1} satisfies Sufficient Condition 1 in Theorem~\ref{thm:M-TDMA},
any M-TDMA policy of any flow decomposition set (e.g., \eqref{equ:ex-L-D}) achieves zero packet loss for all TS flows.
However, since $\ost_{i,j} \ge 0$ for all flow $f_{i,j}$, for any matching $\bm{M}_k$ of any flow decomposition $\mathcal{D}$,
the resulting $T_k$ is at least 2 according to \eqref{equ:t1} and \eqref{equ:t2}, and thus $\sum_{k=1}^4 \frac{1}{T_k} \ge 2$.
Hence, we cannot find a $T$-vector to meet \eqref{equ:condition-basis} for all flow decomposition sets. Therefore, Sufficient
Condition 2 does not hold for this example.
\end{example}

\begin{example} \label{ex2:EDF}
Consider a $4 \times 4$ switch with TS traffic whose offset matrix and period matrix are respectively set as follows,
\bee
\resizebox{.88\hsize}{!}{$
\bm{\ost}=
\left(
  \begin{array}{cccc}
    0 & 0 & 0 & 0 \\
    0 & 0 & 0 & 0 \\
    0 & 0 & 0 & 0 \\
    0 & 0 & 0 & 0 \\
  \end{array}
\right),
\bm{T}=
\left(
  \begin{array}{cccc}
    2 & 4 & 8 & 8 \\
    8 & 2 & 4 & 8 \\
    8 & 8 & 2 & 4 \\
    4 & 8 & 8 & 2 \\
  \end{array}
\right).
$}
\label{equ:ex2}
\eee
We can see that this example cannot satisfy Sufficient Condition 1 since there exists $T_{i,j} < N=4$. However, we can find the following flow decomposition set in its corresponding Latin square,
\be
\bm{L}=
\left(
  \begin{matrix}
    \bm{M}_1 & \bm{M}_2 & \bm{M}_3 & \bm{M}_4 \\
    \bm{M}_4 & \bm{M}_1 & \bm{M}_2 & \bm{M}_3 \\
    \bm{M}_3 & \bm{M}_4 & \bm{M}_1 & \bm{M}_2 \\
    \bm{M}_2 & \bm{M}_3 & \bm{M}_4 & \bm{M}_1 \\
  \end{matrix}
\right),
\ee
and its resulting $T$-vector $(2, 4, 8, 8)$. We can check that
either condition \eqref{equ:thm-M-EDF-con1} or condition \eqref{equ:thm-M-EDF-con2} holds for any flow $f_{i,j} \in \bm{M}_k$.
Thus, this example satisfies Sufficient Condition 2 and the corresponding M-EDF policy achieves zero packet loss for all TS flows.
In fact, the scheduled matching at each slot is shown in Appendix C.

%Appendix~\ref{app:an-example-EDF}.
\end{example}

We remark that our independent simulations confirm that M-TDMA (resp. M-EDF) indeed achieves zero packet loss for all TS flows
in Example~\ref{ex1:TDMA} (resp. Example~\ref{ex2:EDF}).

In addition, we also illustrate an example to schedule both TS and BE flows in
Appendix D.
%\ifx \ISTR \undefined
%Appendix \ref{app:schedule-both-TS-and-BE-flows} in our technique report \cite{TR}.
%\else
%Appendix \ref{app:schedule-both-TS-and-BE-flows}.
%\fi
%Appendix~\ref{app:schedule-both-TS-and-BE-flows} in our technique report \cite{TR}.

\section{Conclusion}
Achieving zero packet loss for timely traffic is an important but challenging requirement in TSN applications.
In this paper, for the first time, we propose an input-queueing TSN switching architecture to achieve this goal.
Specifically, we propose two sufficient conditions and two corresponding scheduling policies (called M-TDMA and M-EDF)
to achieve zero packet loss for all timely traffics. These two sufficient conditions have non-empty intersections, and
no one is more strict than the other. Thus, both conditions are not necessary for achieving zero packet loss for timely traffic.
It is very interesting and important to characterize a sufficient and necessary condition to achieve zero packet loss for timely traffic
in the future.

\bibliographystyle{IEEEtran}
\bibliography{ref}

%\newpage

\ifx \ISTR \undefined
\else
\appendix

\subsection{An Example of Flow Decomposition Set and Its Corresponding Latin Square} \label{app:ex-flow-decomposition-set}
Consider a $4 \times 4$ switch, i.e., $N=4$. We construct the following 4 perfect matchings,
\bee
&& \bm{M}_1 =
\left(
  \begin{matrix}
    1 & 0 & 0 & 0 \\
    0 & 1 & 0 & 0 \\
    0 & 0 & 1 & 0 \\
    0 & 0 & 0 & 1 \\
  \end{matrix}
\right),
\bm{M}_2 =
\left(
  \begin{matrix}
    0 & 1 & 0 & 0 \\
    0 & 0 & 1 & 0 \\
    0 & 0 & 0 & 1 \\
    1 & 0 & 0 & 0 \\
  \end{matrix}
\right),  \nnb \\
&& \bm{M}_3 =
\left(
  \begin{matrix}
    0 & 0 & 1 & 0 \\
    0 & 0 & 0 & 1 \\
    1 & 0 & 0 & 0 \\
    0 & 1 & 0 & 0 \\
  \end{matrix}
\right),
\bm{M}_4 =
\left(
  \begin{matrix}
    0 & 0 & 0 & 1 \\
    1 & 0 & 0 & 0 \\
    0 & 1 & 0 & 0 \\
    0 & 0 & 1 & 0 \\
  \end{matrix}
\right).
\label{equ:app-M1-M4}
\eee
We can examine that
\be
\bm{M}_1 + \bm{M}_2 + \bm{M}_3 + \bm{M}_4 =
\left(
  \begin{matrix}
    1 & 1 & 1 & 1 \\
    1 & 1 & 1 & 1 \\
    1 & 1 & 1 & 1 \\
    1 & 1 & 1 & 1 \\
  \end{matrix}
\right),  \label{equ:app-sum-M}
\ee
where the addition is operated in the real field.
Thus, $\mathcal{D}=\{\bm{M}_1, \bm{M}_2, \bm{M}_3, \bm{M}_4\}$ is a flow decomposition set.

Now we can construct its Latin square whose entry in the $i$-th row and $j$-th column
is the perfect matching in  $\mathcal{D}$ containing flow $f_{i,j}$, i.e.,
\bee
\bm{L}(\mathcal{D})  =
\left(
  \begin{matrix}
    \bm{M}_1 & \bm{M}_2 & \bm{M}_3 & \bm{M}_4 \\
    \bm{M}_4 & \bm{M}_1 & \bm{M}_2 & \bm{M}_3 \\
    \bm{M}_3 & \bm{M}_4 & \bm{M}_1 & \bm{M}_2 \\
    \bm{M}_2 & \bm{M}_3 & \bm{M}_4 & \bm{M}_1 \\
  \end{matrix}
\right).
\label{equ:app-D-matrix}
\eee
Clearly, the constructed Latin square is unique because any flow belongs to exactly one perfect matching
in a flow decomposition set.

In addition, since $l_{i,j}$ is the perfect matching containing flow $f_{i,j}$, we can easily see that
all same entries of $\bm{L}(\mathcal{D})$ in \eqref{equ:app-D-matrix} reconstruct the corresponding perfect matching.
For example, in \eqref{equ:app-D-matrix}, we can see that $l_{1,1}=l_{2,2}=l_{3,3}=l_{4,4}=\bm{M}_1$, and thus we can reconstruct $\tilde{\bm{M}}_1$ as
\be
\tilde{\bm{M}}_1 =
\left(
  \begin{matrix}
    1 & 0 & 0 & 0 \\
    0 & 1 & 0 & 0 \\
    0 & 0 & 1 & 0 \\
    0 & 0 & 0 & 1 \\
  \end{matrix}
\right),  \label{equ:app-M1-reconstruct}
\ee
which is the same as $\bm{M}_1$ in \eqref{equ:app-M1-M4}. This reconstruction step will be used to prove Theorem~\ref{the:iff-conditon} in Appendix~\ref{app:proof-of-the-iff-condition}.

\subsection{Proof of Theorem~\ref{the:iff-conditon}} \label{app:proof-of-the-iff-condition}
First, according the construction procedure in Sec.~\ref{subsec:flow-decomp-set},
it is easily to see that for any flow decomposition set $\mathcal{D}$, its corresponding Latin square $\bm{L}(\mathcal{D})$ is
unique and the first row is $(\bm{M}_1, \bm{M}_2, \cdots, \bm{M}_N)$.

To prove Theorem~\ref{the:iff-conditon}, we now only need to show that for any
Latin square $\tilde{\bm{L}}$ of order $N$ whose first row is fixed as $(\bm{M}_1,\bm{M}_2, \cdots, \bm{M}_N)$, there exists a unique
flow decomposition set $\mathcal{D}$ such that
\be
\bm{L}(\mathcal{D}) = \tilde{\bm{L}}. \label{equ:app-L-D=L}
\ee

We first construct a flow decomposition set $\tilde{\mathcal{D}}$
such that \eqref{equ:app-L-D=L} holds. This can be done as follows.
We construct an $N\times N$ matrix $\tilde{\bm{M}}_k=(\tilde{m}_{i,j} \in \{0,1\}: i,j=1,2,\cdots,N)$ for any $k \in [N]$
where
\be
\tilde{m}_{i,j}=
\left\{
  \begin{array}{ll}
    1, & \hbox{If $ \tilde{l}_{i,j}=\bm{M}_k$;} \\
    0, & \hbox{Otherwise.}
  \end{array}
\right.
\ee
Namely, we extract all entries whose value is $\bm{M}_k$ in Latin square $\tilde{\bm{L}}$ to construct the index matrix $\tilde{\bm{M}}_k$ (see an example in \eqref{equ:app-M1-reconstruct}).
Since $\tilde{\bm{L}}$ is a Latin square whose any row or column does not have the same entry, we can see that
$\tilde{\bm{M}}_k$ is a perfect matching. Then we construct the flow decomposition set $\mathcal{\tilde{D}}=\{\tilde{\bm{M}}_1,\tilde{\bm{M}}_2,\cdots,\tilde{\bm{M}}_N\}$.
According the construction procedure in Sec.~\ref{subsec:flow-decomp-set}, we can see that $\bm{L}(\mathcal{\tilde{D}}) = \tilde{\bm{L}}$. Namely,
there exists at least one flow decomposition set $\mathcal{\tilde{D}}$ such that \eqref{equ:app-L-D=L} holds.

Now we prove that for any two different flow decomposition sets $\mathcal{D}'$ and $\mathcal{D}''$, their constructed Latin squares must be different, i.e.,
\be
\bm{L}(\mathcal{D}') \neq \bm{L}(\mathcal{D}''). \label{equ:app-L-D1-neq-L-D2}
\ee
Let us denote
\be
\bm{L}' \triangleq \bm{L}(\mathcal{D}'), \quad \bm{L}'' \triangleq \bm{L}(\mathcal{D}'').
\ee

Since flow decomposition sets $\mathcal{D}'$ and $\mathcal{D}''$ are different, there must exist a flow $f_{i,j}$ such that
the matching containing flow $f_{i,j}$ in $\mathcal{D}'$, denoted as $\bm{M}'$, is different from the matching containing flow $f_{i,j}$ in $\mathcal{D}''$, denoted as $\bm{M}''$, i.e.,
\be
\bm{M}' \neq \bm{M}''.
\ee

According the construction procedure in Sec.~\ref{subsec:flow-decomp-set}, the entry in the $i$-th row and $j$-th column
of the constructed Latin square is the perfect matching containing flow $f_{i,j}$. Thus,
\be
l_{i,j}'=\bm{M}', \quad l_{i,j}''=\bm{M}''.
\ee

Therefore, $\bm{L}' \neq \bm{L}''$ and thus \eqref{equ:app-L-D1-neq-L-D2} holds. Hence, there exists a unique
flow decomposition set $\mathcal{D}$ such that \eqref{equ:app-L-D=L} holds.

The proof is thus completed.

\subsection{An Example of Virtual Single-processor Task-execution System under EDF Scheduling Policy} \label{app:an-example-EDF}
We consider $N=4$ and  a $T$-vector
\be
\overrightarrow{T}=(T_1, T_2, T_3, T_4) = (2,4,8,8). \label{equ:app-ex-EDF-T-vec}
\ee
Note that we can check that \eqref{equ:condition-basis} holds and thus
\eqref{equ:app-ex-EDF-T-vec} is indeed a $T$-vector. Based on this $T$-vector $\overrightarrow{T}$,
we construct a virtual single-processor task-execution system which has four tasks with periods specified by  $\overrightarrow{T}$.
We illustrate the arrived requests of all four tasks in the virtual single-processor system
in Fig.~\ref{fig:app-EDF-example}.

At each slot, EDF scheduling policy  schedules the request of all $N$ tasks
who has the earliest deadline and breaks ties arbitrarily.
Then, we can construct the index of the scheduled task at any slot $t$, i.e., $a^{\textsf{EDF}}_t(\overrightarrow{T}) \in [N]$,
as shown in the bottom of Fig.~\ref{fig:app-EDF-example}.

\begin{figure}[t]
  \centering
  \includegraphics[width=\linewidth]{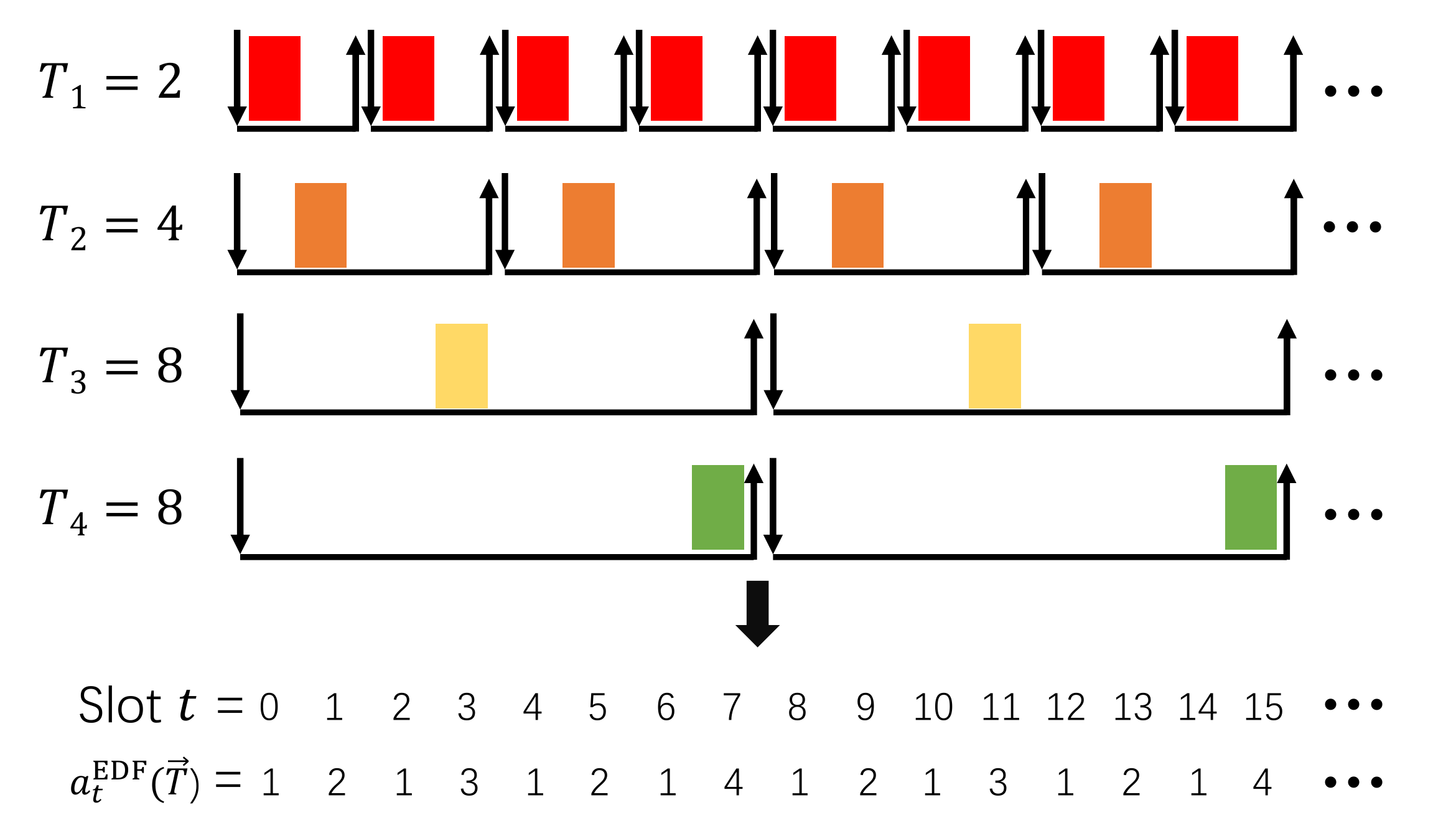}\\
  \caption{An example of the virtual single-processor task-execution system.}\label{fig:app-EDF-example}
\end{figure}

\subsection{An Example to Schedule both TS Flows and BE Flows} \label{app:schedule-both-TS-and-BE-flows}
In the main parts of this paper, we consider that only TS flows exist in the system. However,
we remark that our proposed TSN switching architecture and scheduling polices can co-schedule both TS flows and BE flows,
as shown in Fig.~\ref{fig:switch}. We give higher priorities to TS flows than BE flows. If the M-TDMA or M-EDF policy cannot
fully utilize the switching ports to schedule TS flows in Step 1, the iSLIP policy will schedule BE flows by utilizing the rest of available ports
in Step 2. Here we use a simple example to illustrate this two-step scheduler of Fig.~\ref{fig:switch}.

We again consider a $4 \times 4$ switch but with only TS flows and BE flows related to input 1. Specifically, there are three TS flows $f_{1,1}, f_{1,2}, f_{1,3}$
with offset and period matrices,
\bee
\resizebox{.88\hsize}{!}{$
\bm{\ost}=
\left(
  \begin{array}{cccc}
    0 & 0 & 0 & \infty \\
    \infty & \infty & \infty & \infty \\
    \infty & \infty & \infty & \infty \\
    \infty & \infty & \infty & \infty \\
  \end{array}
\right),
\bm{T}=
\left(
  \begin{array}{cccc}
    3 & 6 & 6 & \infty \\
    \infty & \infty & \infty & \infty \\
    \infty & \infty & \infty & \infty \\
    \infty & \infty & \infty & \infty \\
  \end{array}
\right).
$}
\label{equ:app-ex-both}
\eee
In addition, we have four BE flows $f^{\textsf{BE}}_{1,1}, f^{\textsf{BE}}_{1,2}, f^{\textsf{BE}}_{1,3}, f^{\textsf{BE}}_{1,4}$, whose
arrivals can be arbitrary since no cell has a strict deadline. We show a collection of arrivals of these four BE flows in Fig.~\ref{fig:app-EDF-example-both}.

\begin{figure}[t]
  \centering
  \includegraphics[width=\linewidth]{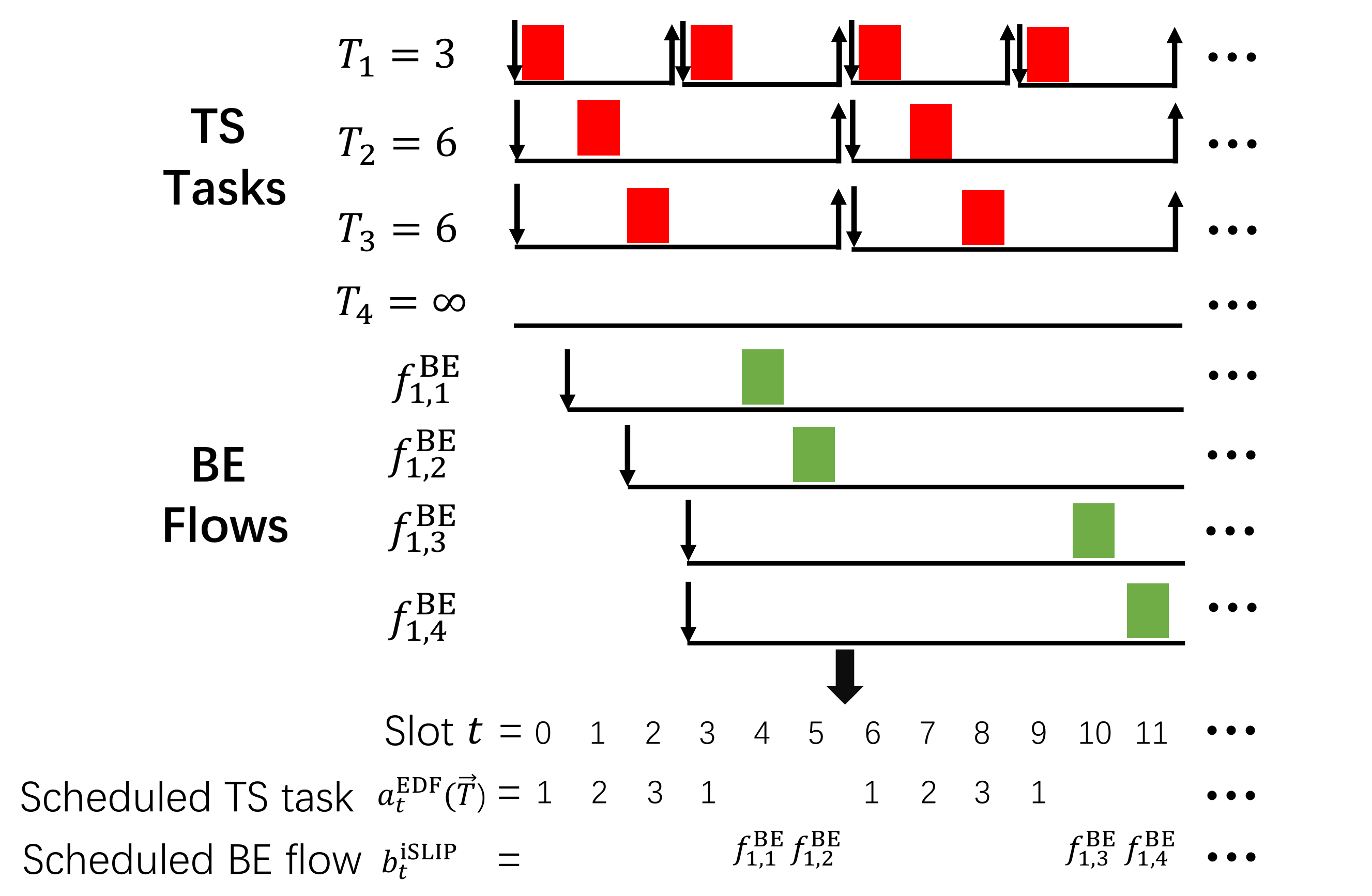}\\
  \caption{An example to schedule both TS flows and BE flows.}\label{fig:app-EDF-example-both}
\end{figure}

We can see that \eqref{equ:app-ex-both} does not satisfy SC1 but satisfy SC2 by constructing the following flow decomposition set (in its Latin-square form),
\be
\bm{L}=
\left(
  \begin{matrix}
    \bm{M}_1 & \bm{M}_2 & \bm{M}_3 & \bm{M}_4 \\
    \bm{M}_4 & \bm{M}_1 & \bm{M}_2 & \bm{M}_3 \\
    \bm{M}_3 & \bm{M}_4 & \bm{M}_1 & \bm{M}_2 \\
    \bm{M}_2 & \bm{M}_3 & \bm{M}_4 & \bm{M}_1 \\
  \end{matrix}
\right),
\ee
and the $T$-vector, $\overrightarrow{T}=(T_1,T_2,T_3,T_4)=(3,6,6,\infty).$
Clearly, we have that TS flows $f_{1,1} \in \bm{M}_1$, $f_{1,2} \in \bm{M}_2$, and $f_{1,3} \in \bm{M}_3$.

At any slot $t$, we first use the M-EDF policy to  schedule TS flows and then use the iSLIP policy to schedule BE flows.
The index of the scheduled matching by the M-EDF policy at any slot $t$, i.e., $a^{\textsf{EDF}}_t(\overrightarrow{T}) \in [N]$,
and the scheduled BE flow at any slot $t$, denoted as $b^{\textsf{iSLIP}}_t \in \{f^{\textsf{BE}}_{1,1}, f^{\textsf{BE}}_{1,2}, f^{\textsf{BE}}_{1,3}, f^{\textsf{BE}}_{1,4}\} $, are shown
in the last two lines of Fig.~\ref{fig:app-EDF-example-both}, respectively.

For example, at slot 4, since all three TS flows do not have any cell in the system,
$a^{\textsf{EDF}}_4$ can be set arbitrarily and no TS flow will be scheduled at this slot.
Therefore, input 1 is not utilized by M-EDF and thus can be utilized by iSLIP to schedule BE flows.
Here, we have one cell of BE flow $f^{\textsf{BE}}_{1,1}$, one cell of BE flow $f^{\textsf{BE}}_{1,2}$,
one cell of BE flow $f^{\textsf{BE}}_{1,3}$, and one cell of BE flow $f^{\textsf{BE}}_{1,4}$ in the system.
The iSLIP policy selects $b^{\textsf{iSLIP}}_4=f^{\textsf{BE}}_{1,1}$ at slot 4.
Later at slot 5, again no TS cell exists in the system and iSLIP schedules $b^{\textsf{iSLIP}}_5=f^{\textsf{BE}}_{1,2}$.

\subsection{Proof of Theorem~\ref{thm:M-TDMA}} \label{app:proof-of-thm-M-TDMA}
Any flow $f_{i,j}$ is contained in one of perfect matchings in $\mathcal{D}=\{\bm{M}_1,\bm{M}_2,\cdots,\bm{M}_N\}$, say $\bm{M}_k$.
According to the M-TDMA policy, perfect matching $\bm{M}_k$ is scheduled at slots $qN+(k-1), q=0,1,\cdots$. Since $T_{i,j} \ge N$,
the $s$-th $(s=0,1,2,\cdots)$ cell (called cell $s$) of flow $f_{i,j}$ arrives at the beginning of slot $\ost_{i,j}+sT_{i,j}$ and expires at the end of slot $\ost_{i,j}+(s+1)T_{i,j}-1$.
Namely, the lifetime of cell $s$ of flow $f_{i,j}$ is the interval $[\ost_{i,j}+sT_{i,j},\ost_{i,j}+(s+1)T_{i,j}-1]$.
Now we set the index of a scheduling period to be
\be
q(s) \triangleq \left \lceil \frac{sT_{i,j}+\ost_{i,j}-(k-1)}{N} \right \rceil.
\ee
Since $s \ge 0, \ost_{i,j} \ge 0, k \le N$, we can derive that $q(s) \ge 0$.
Since $\bm{M}_k$ is scheduled at slot $q(s)N+(k-1)$ in scheduling period $q(s)$,
to ensure that cell $s$ of flow $f_{i,j}$ gets a scheduling slot,
we only need to show that
\be
\ost_{i,j}+sT_{i,j} \le q(s) N + (k-1) \le \ost_{i,j}+(s+1)T_{i,j}-1,
\ee
i.e.,
\be
\frac{sT_{i,j}+\ost_{i,j}-(k-1)}{N} \le q(s) \le  \frac{(s+1)T_{i,j}+\ost_{i,j}-k}{N}
\label{equ:q(s)-inequ}
\ee
If $q(s) =  \frac{sT_{i,j}+\ost_{i,j}-(k-1)}{N} $, then \eqref{equ:q(s)-inequ} holds directly.
Otherwise, if $q(s) >  \frac{sT_{i,j}+\ost_{i,j}-(k-1)}{N} $, then
\bee
q(s) & = \left \lceil \frac{sT_{i,j}+\ost_{i,j}-(k-1)}{N} \right \rceil \nnb \\
& < \frac{sT_{i,j}+\ost_{i,j}-(k-1)}{N} + 1.
\eee
Thus,
\be
q(s)N < sT_{i,j}+\ost_{i,j}-(k-1) + N. \label{equ:qsN-inequ}
\ee
Since both sides of \eqref{equ:qsN-inequ} are integers, we have
\be
q(s)N \le sT_{i,j}+\ost_{i,j}-(k-1) + N - 1.
\ee
Since $T_{i,j}\ge N$, we thus have,
\bee
q(s) & \le  \frac{sT_{i,j}+\ost_{i,j}-(k-1) + N - 1}{N} \nnb \\
& \le \frac{sT_{i,j}+\ost_{i,j}-(k-1) + T_{i,j} - 1}{N} \nnb \\
& = \frac{(s+1)T_{i,j}+\ost_{i,j}-k}{N}.
\eee
Thus, \eqref{equ:q(s)-inequ} holds. Therefore,
cell $s$ of flow $f_{i,j}$ will be scheduled at slot $q(s)N+(k-1)$ before expiration.
Since $s$ is arbitrary, any cell of flow $f_{i,j}$ will be allocated one scheduling slot.
In addition, since $T_{i,j}\ge N$, it is easy to see that $q(s) \neq q(s')$ if $s \neq s'$.
Thus, any two cells of flow $f_{i,j}$ will be not be scheduled at the same slot.
We thus know that any cell will be allocated one unique scheduling slot before expiration.

Therefore, M-TDMA achieves zero packet loss for all flows.

\subsection{Proof of Theorem~\ref{thm:M-EDF}} \label{app:proof-of-thm-M-EDF}
Any flow $f_{i,j}$ is contained in one of perfect matchings in $\mathcal{D}=\{\bm{M}_1,\bm{M}_2,\cdots,\bm{M}_N\}$, say $\bm{M}_k$.
According to the M-TDMA policy, perfect matching $\bm{M}_k$ is scheduled at slots $qN+(k-1), q=0,1,\cdots$. Since $T_{i,j} \ge N$,
the $s$-th $(s=0,1,2,\cdots)$ cell (called cell $s$) of flow $f_{i,j}$ arrives at the beginning of slot $\ost_{i,j}+sT_{i,j}$ and expires at the end of slot $\ost_{i,j}+(s+1)T_{i,j}-1$. We consider the $s$-th cell (called cell $s$) of the flow $f_{i,j} \in \bm{M}_k$,
whose lifetime is time interval $[\ost_{i,j}+sT_{i,j},\ost_{i,j}+(s+1)T_{i,j}-1]$.

\textbf{Case 1.} If condition \eqref{equ:thm-M-EDF-con1} holds,
the lifetime of cell $s$ of flow $f_{i,j}$ is $[sT_k, (s+1)T_k-1]$, which is exactly the lifetime of the $s$-th request (called request $s$) of task $k$ in the
virtual single-processor task scheduling system. Since EDF achieves zero packet loss for any $T$-vector,
request $s$ of task $T_k$ is executed at some slot $t' \in [sT_k, (s+1)T_k-1]$, i.e., $a^{\textsf{EDF}}_{t'}(\overrightarrow{T})=k$.
Since the M-EDF policy schedules matching $\bm{M}_{a^{\textsf{EDF}}_t(\overrightarrow{T})}$ at any slot $t$, it schedules
matching  $\bm{M}_{a^{\textsf{EDF}}_{t'}(\overrightarrow{T})}=\bm{M}_k$, which contains flow $f_{i,j}$, at slot $t'$.
Therefore, cell $s$ of flow $f_{i,j}$ must be scheduled within its lifetime.

\textbf{Case 2.} If condition \eqref{equ:thm-M-EDF-con2} holds, we need to find a slot $t' \in [\ost_{i,j}+sT_{i,j},\ost_{i,j}+(s+1)T_{i,j}-1]$
such that
\be
a^{\textsf{EDF}}_{t'}(\overrightarrow{T})=k.
\ee
Namely, we should prove that the EDF policy  schedules a request of task $k$ at some slot
$t' \in [\ost_{i,j}+sT_{i,j},\ost_{i,j}+(s+1)T_{i,j}-1]$ in the {virtual} single-processor task scheduling system.
Define
\be
q(s) \triangleq \left \lceil  \frac{sT_{i,j}+\ost_{i,j}}{T_k} \right \rceil.
\ee
Note that $q(s) T_k$ is the arrival time of request $q(s)$ of task $T_k$ in the
{virtual} single-processor task scheduling system. Now we first prove that $q(s) T_k$
is within the lifetime of cell $s$ of flow $f_{i,j}$ in the switch system.
We can see that
\bee
q(s) T_k & = \left \lceil  \frac{sT_{i,j}+\ost_{i,j}}{T_k} \right \rceil \times T_k \nnb \\
& <  \left( \frac{sT_{i,j}+\ost_{i,j}}{T_k} + 1 \right) \times T_k \nnb \\
& = sT_{i,j} + T_k + \ost_{i,j} \nnb \\
& \le sT_{i,j} + \frac{T_{i,j}+1}{2} + \ost_{i,j} \nnb \\
& = \ost_{i,j} + (s+1)T_{i,j}-1 + \frac{3-T_{i,j}}{2} \nnb \\
& \le \ost_{i,j} + (s+1)T_{i,j}-1,
\eee
where the last inequality follows from \eqref{equ:thm-M-EDF-con2} and $\overrightarrow{T}$ is a $T$-vector as
$T_{i,j} \ge 2T_k - 1 \ge 2\times 2 -1 =3.$

Thus, we have
\be
q(s) T_k \in [\ost_{i,j}+sT_{i,j},\ost_{i,j}+(s+1)T_{i,j}-1]. \nnb
\ee
Therefore, request $q(s)$ of task $T_k$ arrives at the virtual system  within
$[\ost_{i,j}+sT_{i,j},\ost_{i,j}+(s+1)T_{i,j}-1]$.
In addition, since EDF achieves zero packet loss with respect to any
$T$-vector, this request must be scheduled at some slot $t'$ before its expiration, i.e.,
\be
t' \le [q(s)+1] T_k -1. \label{equ:the-EDF-t-expiration}
\ee
Now let us prove
\be
[q(s)+1] T_k -1 \le  \ost_{i,j}+(s+1)T_{i,j}-1, \nnb
\ee
which is equivalent to
\be
\left \lceil  \frac{sT_{i,j}+\ost_{i,j}}{T_k} \right \rceil T_k \le \ost_{i,j}+(s+1)T_{i,j} - T_k. \label{equ:the-EDF-q(s)-inequ}
\ee
If $\frac{sT_{i,j}+\ost_{i,j}}{T_k}$ is an integer, then
\bee
& \left \lceil  \frac{sT_{i,j}+\ost_{i,j}}{T_k} \right \rceil T_k = \frac{sT_{i,j}+\ost_{i,j}}{T_k} \cdot T_k \nnb \\
& = sT_{i,j}+\ost_{i,j} \le sT_{i,j}+\ost_{i,j} + (T_{i,j} - T_k) \nnb \\
& = \ost_{i,j}+(s+1)T_{i,j} - T_k.
\eee
Thus, \eqref{equ:the-EDF-q(s)-inequ} holds.
Otherwise, if $\frac{sT_{i,j}+\ost_{i,j}}{T_k}$ is not an integer, then
\bee
\left \lceil  \frac{sT_{i,j}+\ost_{i,j}}{T_k} \right \rceil T_k &  < \left( \frac{sT_{i,j}+\ost_{i,j}}{T_k} +1 \right) \times T_k \nnb \\
& = sT_{i,j}+\ost_{i,j} + T_k. \label{equ:the-EDF-key-inequ}
\eee
Since both sides of \eqref{equ:the-EDF-key-inequ} are integers, we have
\bee
& \left \lceil  \frac{sT_{i,j}+\ost_{i,j}}{T_k} \right \rceil T_k \le sT_{i,j}+\ost_{i,j} + T_k - 1 \nnb \\
& \le sT_{i,j}+\ost_{i,j} + (T_{i,j} - T_k + 1) - 1 \nnb \\
& = \ost_{i,j}+(s+1)T_{i,j} - T_k,
\eee
where the last inequality follows from \eqref{equ:thm-M-EDF-con2}.
Thus, \eqref{equ:the-EDF-q(s)-inequ} holds.

Therefore, \eqref{equ:the-EDF-t-expiration} holds. Namely, request $q(s)$ of task $T_k$ is scheduled at slot
$t' \in [\ost_{i,j}+sT_{i,j},\ost_{i,j}+(s+1)T_{i,j}-1]$ in the virtual system.
Then, since the M-EDF policy schedules matching $\bm{M}_{a^{\textsf{EDF}}_t(\overrightarrow{T})}$ at any slot $t$, it will schedule
matching $\bm{M}_{a^{\textsf{EDF}}_{t'}(\overrightarrow{T})}=\bm{M}_k$ containing flow $f_{i,j}$ at slot $t'$.
Therefore, cell $s$ of flow $f_{i,j}$ must be scheduled within its lifetime.

Since $s$ is selected arbitrarily for any flow $f_{i,j} \in \bm{M}_k$,
\textbf{Case 1}  and \textbf{Case 2} finish the proof.

\fi

\end{document}